\documentstyle[12pt,fleqn]{article}
\begin{document}
\begin{titlepage}
\thispagestyle{empty}
\begin{flushright}
\vbox{
{\bf TTP 94-26}\\
{\bf November 1994}\\
}
\end{flushright}
\vspace{0.5cm}
\begin{center}
{\huge Radiative corrections to $b\to c\tau\bar\nu_\tau$}
\end{center}
\vskip1.0cm
\begin{center}
{\large\bf Andrzej Czarnecki}\\
\vskip0.3cm
{\it Institut f\"ur Theoretische Teilchenphysik,
D-76128 Karlsruhe, Germany} \\
\vskip1.0cm
{\large\bf Marek Je\.zabek}\\
\vskip0.3cm
{\it Institute of Nuclear Physics, Kawiory 26a, PL-30055 Cracow,
Poland}\\
{\it Institut f\"ur Theoretische Teilchenphysik,
D-76128 Karlsruhe, Germany} \\
\vskip0.7cm
{\large and}\\
\vskip0.7cm
{\large\bf Johann H. K\"uhn} \\
\vskip0.3cm
{\it Institut f\"ur Theoretische Teilchenphysik,
D-76128 Karlsruhe, Germany}
\end{center}
\vskip1.5cm
\begin{center}
{\large Abstract}
\end{center}
Analytical calculation is presented of the QCD radiative corrections
to the rate of the process $b\to c\tau\bar\nu_\tau$
and to the $\tau$ lepton longitudinal polarization
in $\tau\bar\nu_\tau$ rest frame. The results are given
in the form of one dimensional infrared finite integrals
over the invariant mass of the leptons. We argue that this form
may be optimal for phenomelogical applications due to a possible
breakdown of semilocal hadron-parton duality in decays of heavy
flavours.

\end{titlepage}
The semileptonic decay rate of $B$ mesons is one of
the key ingredients in the determination
of weak mixing angles.
Transitions of the $b$ quark to the charmed
as well as to the up quark have been analysed in
great detail, exploiting either the inclusive decay rate
or exclusive channels.
The analysis of the inclusive rate is, however,
affected by the uncertainty in the $b$ mass
and by bound state corrections.
This problem is only partly circumvented by fixing
$m_b-m_c$ through the difference of bottom and charmed
meson masses and by relating the bound state effects
to phenomenological constants that can be determined from
other observables in the context of the heavy quark effective
theory (HQET). The decay rate is, furthermore, affected by
perturbative QCD corrections, which have been calculated
analytically up to order $\alpha_s$ for arbitrary $b$
and $c$ masses and massless leptons \cite{CM,JK,Nir}.
Numerical results for perturbative QCD corrections
to the partial decay rate $b\to \tau\bar\nu X$
have been recently obtained in Ref.\cite{FLNN}.
The bound state corrections to this decay chanel
up to order ${1/ m_b^2}$ are also known
\cite{FLNN,Koyrakh,BKPS}
in the context of the HQET.
The comparison between
decays into light ($\mu$, $e$) and heavy ($\tau$) leptons may
furthermore help to test the theoretical approach and allow to fix
some of the free parameters. Including $b\to u$ transitions, four
kinematically different leptonic decay modes are thus available for
the comparison.

In this paper analytical results for the decay rate
$b\to c\tau\bar\nu_\tau$
are presented
in the form of an one dimensional integral over the invariant mass
squared ${\rm w}^2$ of the leptonic system.  This formulation allows,
at least in principle, the separation of the region of
relatively small ${\rm w}^2$, where the inclusive parton model
description
based on local parton-hadron duality should work, from the
region of large ${\rm w}^2$ where only one or few resonances
are produced and the duality between the parton
and hadron description
may be doubtful\footnote{In a recent
preprint \cite{BDS} the breakdown
of the local parton-hadron duality
has been invoked
as the origin of problems with the semileptonic decay rate
of $D$ mesons in the framework of HQET. Let us remark that for
large ${\rm w}^2$ the kinetic energy
of the hadronic system in $B$ decays
can be similar to that in $D$ decays. Thus the
semileptonic branching ratios for $b$ decays may be also
affected for the effective mass of the hadronic system
close to the resonance region.}.
In the region of large ${\rm w}^2$ i.e. close
to the Shifman-Voloshin limit\cite{VS}
${\rm w}^2= {\rm w}^2_{max}$
the theoretical description of lepton spectra
based on summation over exclusive channels
is particularly simple and reliable.
The HQET approach for exclusive decays\cite{IW,Neubert}
on the other hand
becomes quite involved if not impractical in the region of
small ${\rm w}^2$ which is dominated by multiparticle final states.

The calculation of the lowest order rate as well
as corrections can be
related in a straightforward way to the corresponding
calculations for
the decay into a virtual $W$ boson with
the subsequent integration
over the mass of the $l\bar\nu$ system.
The differential decay rate is proportional to
$$ {\cal H}^{\alpha\beta}\,{\cal L}_{\alpha\beta}\,
{\rm dPS}(b\to c\tau\bar\nu)$$
${\cal H}_{\alpha\beta}$ depends on quark and gluon fields
and
\begin{eqnarray}
{\cal L}_{\alpha\beta}(\tau ;\nu) &\sim&
\sum_{s}
\left[\,\bar u_\tau\gamma_\alpha(1
-\gamma_5) v_\nu\,\right]\,
\left[\,\bar u_\tau\gamma_\beta (1-\gamma_5)v_\nu\,\right]^\dagger
\nonumber\\
&\sim& \nu_\alpha \tau_\beta +
\tau_\alpha \nu_\beta -  \nu\cdot\tau g_{\alpha\beta}
- {\rm i}\varepsilon_{\alpha\beta\gamma\delta}\nu^\gamma\tau^\delta
\label{eq:Lab}
\end{eqnarray}
where $\tau^\alpha$ and $\nu^\alpha$ are the four-momenta of
$\tau$ and $\bar\nu_\tau$.
The phase space for the decay of $b$ into $c\tau\bar\nu$
is, in the standard way,
decomposed into a sequence of two-particle final
states
\begin{equation}
{\rm dPS}(b\to c\tau\bar\nu)\, \sim\,
{\rm d}{\rm w}^2\, {\rm dPS}(b\to c{\rm w})\,
{\rm dPS}({\rm w}\to \tau\bar\nu)
\end{equation}
where ${\rm w}^\alpha=\tau^\alpha +\nu^\alpha$.
Then, it is straightforward to show that
\begin{eqnarray}
\lefteqn{
\int {\rm dPS}({\rm w}\to \tau\bar\nu){\cal L}_{\alpha\beta}\, \sim
{\cal A}\left(m_\tau^2/{{\rm w}}^2\right)\,T^{(0)}_{\alpha\beta}\,+
{\cal B}\left(m_\tau^2/{{\rm w}}^2\right)\,T^{(1)}_{\alpha\beta} =
}
\nonumber\\
&& {\textstyle
\left(\, 1 - {m_\tau^2/ {{\rm w}}^2}\, \right)^2
\left[\,\left(\,1 + {2m_\tau^2/ {{\rm w}}^2}\,\right)
{\rm w}_\alpha {\rm w}_\beta
- \left(\,{{\rm w}}^2+ {1\over 2} m_\tau^2\, \right)\,g_{\alpha\beta}
\right]
  }
\label{eq:intL}
\end{eqnarray}
where
\begin{eqnarray}
T^{(0)}_{\alpha\beta} &=&
{\rm w}_\alpha {\rm w}_\beta
\nonumber\\
T^{(1)}_{\alpha\beta} &=&
{\rm w}_\alpha {\rm w}_\beta - {\rm w}^2 g_{\alpha\beta}
\end{eqnarray}
It follows that
the decay rate can be split accordingly into two
incoherent terms which are related to weak decays
of a heavy quark $Q$ into  another quark $q$
and a real spin one or a spin zero boson.
The relative weight of spin one versus spin zero
contributions is governed by their respective spectral
functions and it can be derived
from eq.(\ref{eq:intL}).
For massless leptons ${\cal A}(0)=0$ and
${\cal B}(0)=1$, and therefore only the
spin one (transversal) component
($\sim T^{(1)}_{\alpha\beta}$)  contributes.
The result is given in \cite{JK}.
For fixed ${\rm w}^2$ this contribution to
the rate can be obtained from the formula for $t\to bW$,
the top quark decay into $b$ quark and a real $W$
boson \cite{JK}\footnote{Note that the rates of up and down
type quark decays into their respective isospin partners are of course
identical, in contrast to the shapes of the lepton spectra.}.
Multiplying this formula by
${\cal B}\left(m_\tau^2/{\rm w}^2\right)$
one obtains the contribution
of the spin one component for $m_\tau\ne 0$.
The other (longitudinal) contribution ($\sim T^{(1)}_{\alpha\beta}$)
can be in an analogous way related to the (yet unobserved)
decay $t\to bH^+$ where $H^+$ denotes a charged Higgs boson.
Let $Q$ and $q$ denote the four-momenta of the quarks in the
initial and in the final state. For a two-body decay mode
the momentum of the $W$ boson is $W= Q-q$. In Born approximation
\begin{eqnarray}
W_\mu \bar u(q)\,\gamma^\mu(1-\gamma_5)\, u(Q) &=&
\bar u(q)\,[\,(\hat Q - \hat q)(1-\gamma_5)\,]\, u(Q)
\nonumber\\
&=& \bar u(q)\,[\,(M-m)+(M+m)\gamma_5\,]\, u(Q)
\label{eq:redu}
\end{eqnarray}
where the equations of motion
\begin{eqnarray}
(\hat Q -M)\,u(Q)=0
\qquad\quad  {\rm and} \qquad\quad
\bar u(q)\,(\hat q-m)=0
\end{eqnarray}
have been used.  The last line in (\ref{eq:redu}) can be
interpreted as the amplitude of the decay
$Q\to qH$ where $H$ is a spin zero boson whose
coupling to the weak quark current is given by
\begin{equation}
g = (M-m) + (M-m)\gamma_5
\end{equation}
Although not applicable for individual Feynman diagrams,
the same relation holds true for the longitudinal contribution
($\sim W_\mu$) to the decay amplitude
when ${\cal O}(\alpha_s)$ QCD corrections are included \cite{CD1}.
Therefore this contribution to the rate $b\to c\tau\bar\nu_\tau$
can be extracted from a formula describing $t\to bH^+$ which
has been given in \cite{CD}; cf. Model I therein.

\noindent
Let us define now the following dimensionless quantities
$$
\textstyle{
\rho= {m_c^2/ m_b^2}\qquad\quad
\eta = {m_\tau^2/ m_b^2}\qquad\quad
{\rm and} \qquad\quad
t = {{\rm w}^2/m_b^2  }  }
$$
and the kinematic functions
\begin{eqnarray}
p_0(t) &=& (1-t+\rho)/2
\nonumber\\
p_3(t) &=& \sqrt{p_0^2 - \rho}
\nonumber\\
p_\pm(t) &=& p_0 \pm p_3 = 1 - w_\mp(t)
\nonumber\\
Y_p(t) &=& {\textstyle{1\over 2}}\ln\left(p_+/ p_-\right)
=\ln\left(p_+/\sqrt{\rho}\right)
\nonumber\\
Y_w(t) &=& {\textstyle{1\over 2}}\ln\left(w_+/w_-\right)
=\ln\left(w_+/\sqrt{t}\right)
\label{eq:rapid}
\end{eqnarray}
where ${\rm w}^\alpha=\tau^\alpha +\nu^\alpha$ and
${\rm w}^2$ denotes the effective mass of $\tau\bar\nu_\tau$.
The partial rate
of the decay $b\to c\tau\bar\nu_\tau$
is given by
\begin{equation}
\Gamma(b\to c\tau\bar\nu_\tau)\:=\:
\int^{(1-\sqrt{\rho})^2}_\eta
{{\rm d}\Gamma\over {\rm d}t}\,{\rm d}t
\label{eq:Gamtot}
\end{equation}
with the
differential rate
\begin{eqnarray}
\lefteqn{
{{\rm d}\Gamma\over {\rm d}t} = \Gamma_{bc}\,
\left( 1-{\eta \over t}\right)^2 \,  \left\{
\left( 1+{\eta \over 2t}\right)
\left[ {\cal F}_0(t) - {2\alpha_s\over 3 \pi} {\cal F}_1(t)\right]
+ {3\eta \over 2t}
\left[ {\cal F}_0^s(t) - {2\alpha_s\over 3 \pi} {\cal F}_1^s(t)\right]
\right\}   }
\nonumber\\
\label{main} \\
\lefteqn{
\Gamma_{bc} =
{G_F^2 m_b^5 |V_{cb}|^2\over 192 \pi^3}
}
\end{eqnarray}
\begin{eqnarray}
{\cal F}_0(t) &=& 4 p_3\,
\left[ \, (1-\rho)^2 + t(1+\rho) - 2 t^2\, \right]
\\
{\cal F}_0^s(t)
&=& 4p_3\,\left[\, (1-\rho)^2 -t ( 1+\rho)\, \right]
\\
{\cal F}_1(t)&=& {\cal A}_1 \Psi  + {\cal A}_2 Y_w
+ {\cal A}_3 Y_p + {\cal A}_4 p_3 \ln\rho + {\cal A}_5 p_3
\\
{\cal F}_1^s(t)&=& {\cal B}_1 \Psi  + {\cal B}_2 Y_w
+ {\cal B}_3 Y_p + {\cal B}_4 p_3 \ln\rho + {\cal B}_5 p_3
\label{massive}
\end{eqnarray}
\begin{eqnarray}
\Psi &=& 8 \ln (2 p_3) -2\ln t\, +\,
\left[ 2{\rm Li}_2 (w_-) -2{\rm Li}_2 (w_+)
 +4 {\rm Li}_2 ({2p_3/ p_+}) \right.
\nonumber \\ && \left.   \hskip65pt
  -4Y_p \ln({2p_3/ p_+})
-\ln p_- \ln w_+ + \ln p_+ \ln w_- \right]\, 2p_0/p_3
\end{eqnarray}
\begin{eqnarray}
{\cal A}_1 &=&  {\cal F}_0(t)
\nonumber\\
{\cal A}_2 &=&
- 8 (1-\rho) \left[ 1 +t -4 t^2-
\rho (2-t) +\rho^2 \right]
\nonumber\\
{\cal A}_3 &=&
- 2 \left[ 3 + 6 t   -21 t^2  + 12 t^3
-\rho (1+12t+5t^2) + \rho^2(11+2t)- \rho^3 \right]
\nonumber\\
{\cal A}_4 &=&
- 6 \left[ 1  + 3 t  - 4 t^2
-\rho (4-t) + 3 \rho^2 \right]
\nonumber\\
{\cal A}_5 &=&
- 2 \left[ 5+9t-6t^2 -\rho( 22 - 9t) + 5 \rho^2 \right]
\\
{\cal B}_1 &=&  {\cal F}^s_0(t)
\nonumber\\
{\cal B}_2 &=& -8 (1-\rho)
\left[(1-\rho)^2-t (1+\rho)\right]
\nonumber\\
{\cal B}_3 &=& - 4(1 -  \rho)^4/t
         -2  (-1 + 3  \rho + 15 \rho^2  - 5 \rho^3)
         + 8  (1 + \rho) t
          - 6  (1 + \rho) t^2
\nonumber\\
{\cal B}_4 &=&      -4 (1-\rho)^3 /t
           -2 (1-\rho) (1-11 \rho)
           +6 (1 + 3 \rho) t
 \nonumber\\
{\cal B}_5 &=&  -6 (1-3 \rho) (3-\rho)
+18 t (1+\rho)
\end{eqnarray}
The integral in eq.(\ref{eq:Gamtot}) can be easily performed
for the Born contribution. It reads:\\
\begin{eqnarray}
\lefteqn{
\Gamma_0(b\to c\tau\bar\nu_\tau) =
\Gamma_{bc}\,
\left\{\,
24\,\left[\,\eta^2(1-\rho^2)\,{\cal Y}_w
+ \rho^2(1-\eta^2)\,{\cal Y}_p\,\right]\, +\,
\right. }
\nonumber\\  &&
\left.
2{\cal P}_3\,\left(1-7\eta-7\eta^2+\eta^3-7\rho
+12\eta\rho -7\eta^2\rho
-7\rho^2 -7\eta\rho^2+\rho^3 \right)
\,\right\}
\label{eq:Gam0tot}
\end{eqnarray}
where
$$
{\cal P}_3 = p_3(\eta),\qquad\quad
{\cal Y}_p = Y_p(\eta)\qquad\quad {\rm and} \qquad\quad
{\cal Y}_w = Y_w(\eta).$$
In principle the first order QCD correction can be also
expressed in terms of polylogarithms. In particular for
$\eta=0$ the formula (12) of \cite{Nir} is obtained.
However, the complete result is lenghty. From the practical
point of view it is much simpler to evaluate the integral
in eq.(\ref{eq:Gamtot}) numerically. Moreover, as it has been
explained, for $t$ close to the Shifman-Voloshin
limit the exclusive description is preferable, thus,
for applications, it may be better to perform this integral
only over a part of the available phase space. In recent
articles \cite{SV,LSW} the size of ${\cal O}(\alpha_s^2)$
corrections has been estimated using the scheme of Brodsky,
Lepage and Mackenzie \cite{BLM}
for fixing the scale $\mu$ of $\alpha_s$. It has turned out
that this scale is rather low. This suggests that
next-to-leading QCD corrections are large. The
most serious problems arise in the region close to the
no-recoil $t=t_{max}$ point.
It is well known, cf. \cite{ACCMM,JK1}, that at
the boundaries of the available phase space logarithmic
divergences may appear as remnants of infrared divergences.
This phenomenon arises when real radiation becomes
supressed relative to virtual corrections. This is exactly
what happens for $t$ in the neighbour of $t_{max}$ and
once again one is led to the conclusion that the inclusive
parton-like description may break down there.

\noindent
In the massless limit $\rho\to 0$
which corresponds to $b\to u\tau\bar\nu_\tau$ transition
the functions in eq.(\ref{massive}) simplify considerably
\begin{eqnarray}
{\cal F}_0(t) &=& 2 (1-t)^2 (1+2t)
\\
{\cal F}_0^s(t)
&=& 2(1-t)^2
\\
{\cal F}_1(t) &=&
{\cal F}_0(t)\,
{\textstyle
\left[\,
{2\over3}\pi^2
+4{\rm Li}_2(t)+2\ln t\ln(1-t)\,\right]
}
      - (1-t)(5+9t - 6t^2)
\nonumber\\  &&
      +\, 4t(1-t-2t^2)\ln t
      + 2(1-t)^2(5+4t)\ln(1-t)
\label{eq:F1y0}
\\
{\cal F}_1^s(t)&=&
\textstyle{
{\cal F}_0^s(t)\,
\left[{2\over 3}\pi^2
+ 4{\rm Li}_2(t) -{9\over2}
+\ln(1-t)\left(2\ln t-{2\over t}+5\right) \right]
}
 + 4(1-t)t\ln t
\nonumber\\  &&
\label{eq:F1sy0}
\end{eqnarray}
After integration over $t$ one derives the following
expression for the total partial rate of $b\to u\tau\bar\nu_\tau$
\begin{eqnarray}
\lefteqn{{1\over \Gamma_{bu}}\Gamma(b\to u\tau\bar \nu_\tau)=
1-8\eta+8\eta^3-\eta^4-12\eta^2\ln\eta}
\nonumber\\&&
-{2\alpha_s\over 3\pi}\left\{
(-1 + \eta) (75 - 539 \eta - 476 \eta^2 + 18 \eta^3)/12
\right.
\nonumber\\&&  \qquad \quad
+ (3 - 24 \eta - 36 \eta^2 + 16 \eta^3 - 2 \eta^4) \pi^2/3
\nonumber\\&&  \qquad  \quad
+ 72 \eta^2 \left[\zeta(3)-{\rm Li}_3(\eta)\right]
 + 2 (1 - 8 \eta + 36 \eta^2 + 16 \eta^3 - 2 \eta^4) {\rm Li}_2(\eta)
\nonumber\\&&  \qquad  \quad
+      (1 - \eta^2) (31 - 320 \eta + 31 \eta^2) \ln(1 - \eta)/6
\nonumber\\&&  \qquad  \quad
 +       \left[2 \eta + 15 \eta^2 - 94 \eta^3/3 + 31 \eta^4/6
          - 8 \eta^2 \pi^2  +  24 \eta^2 {\rm Li}_2(\eta)
\right.        \qquad  \quad
\nonumber\\&& \left.\left.  \qquad \qquad
+   2 (1 - \eta^2) (1 - 8 \eta +\eta^2) \ln(1 - \eta)\right] \ln(\eta)
\right\}
\end{eqnarray}
\begin{eqnarray}
\lefteqn{
\Gamma_{bu} =
{G_F^2 m_b^5 |V_{ub}|^2\over 192 \pi^3} }
\end{eqnarray}
which agrees with eqs.(2.21) and (3.6)
in a recent preprint \cite{BBBG}.

It has been argued \cite{kalinowski} that the $\tau$ polarization
in $b\to c\tau\bar\nu_\tau$ is particularly sensitive to deviations
from the Standard Model. The longitudinal component of
$\tau$ polarization can either be measured
with reference to its
direction of flight in the $b$ rest frame
or, alternatively, relative
to its direction of flight in the $\tau\bar\nu$ rest frame.
To evaluate analytically
QCD corrections to the longitudinal $\tau$ polarization
in the $b$ rest frame is a demanding task.
Previous experience from a similar calculation
for the decay of polarized top quarks indicates
that these corrections are typically quite small through most of the
kinematical range.  To substantiate this claim,
the longitudinal $\tau$ polarization in the $\tau\bar\nu$ rest
frame is evaluated including QCD corrections.

\noindent
For V-A weak current the leptonic tensor
\begin{eqnarray}
{\cal L}_{\alpha\beta}(\tau,s;\nu) &\sim&
\left[\,\bar u_\tau\gamma_\alpha(1
-\gamma_5) v_\nu\,\right]\,
\left[\,\bar u_\tau\gamma_\beta (1-\gamma_5)v_\nu\,\right]^\dagger
\nonumber\\
&\sim& \nu_\alpha {\cal T}_\beta +
{\cal T}_\alpha \nu_\beta -  \nu\cdot{\cal T} g_{\alpha\beta}
- {\rm i}\varepsilon_{\alpha\beta\gamma\delta}\nu^\gamma
{\cal T}^\delta
\label{eq:Labs}
\end{eqnarray}
where
\begin{equation}
{\cal T}^\alpha = {\textstyle{1\over 2}}
\left(\,\tau^\alpha -m_\tau s^\alpha\,\right)
\end{equation}
and $s^\alpha$ is the $\tau$ polarization fourvector.
The helicity states in the $\tau\bar\nu$
rest frame are obtained for
\begin{equation}
m_\tau s^\alpha = \pm \left( \tau^\alpha
- {\textstyle{2\eta\over t-\eta}} \nu^\alpha \right)
\end{equation}
It is evident that the net polarization can be calculated
in the same way as the total rate, decomposing again the
spin dependent term into longitudinal and transversal parts.
For the positive helicity of $\tau$ one derives the following
expression for the differential rate
\begin{eqnarray}
\lefteqn{
{1\over \Gamma_{bc}}\,{{\rm d}\Gamma^{(+)}\over {\rm d}t} =
{\eta\over 2t}
\left( 1-{\eta \over t}\right)^2 \,  \left\{
{\cal F}_0(t) + 3 {\cal F}_0^s(t)
- {2\alpha_s\over 3 \pi}
\left[\,
{\cal F}_1(t) + 3 {\cal F}_1^s(t)\,\right]
\right\}   }
\label{polar}
\end{eqnarray}
For the negative helicity one has
\begin{eqnarray}
\lefteqn{
{{\rm d}\Gamma^{(-)}\over {\rm d}t} =
{{\rm d}\Gamma\over {\rm d}t} -
{{\rm d}\Gamma^{(+)}\over {\rm d}t}
  }
\label{polar1}
\end{eqnarray}
and the net helicity in the
$\tau\nu_\tau$ rest frame is equal to
\begin{equation}
{\rm P} = - 1 \, + \,
{ 2
\int^{(1-\sqrt{\rho})^2}_\eta
{{\rm d}\Gamma^{(+)}\over {\rm d}t}\,{\rm d}t
\over
\int^{(1-\sqrt{\rho})^2}_\eta
{{\rm d}\Gamma\over {\rm d}t}\,{\rm d}t
}
\end{equation}

In the subsequent discussion the mass difference $m_b-m_c$
will be fixed through the HQET relation
$$m_b-m_c = [ (3m_{B^*}+m_{B})-(3m_{D^*}+m_{D})]/4 + \ldots$$
The most important corrections to the above relation arise from
the kinetic energy of heavy quarks in $B$ and $D$ mesons. On
physical grounds one expects this contribution to increase
the difference between $m_b$ and $m_c$.
We take
$m_b- m_c=$~3.4~GeV
in numerical calculations
and $m_b$ is varied between
4.5 and 5.0~GeV. In order to estimate effects of QCD corrections
on measurable quantities we neglects all the problems with
the parton-hadron duality and the scale ambiguity. The following
results have been obtained:
\begin{eqnarray}
&&\Gamma(b\to c\tau\bar\nu_\tau) =
\Gamma_{0}(b\to c\tau\bar\nu_\tau) \,
\left[\, 1 + ( -0.450\pm 0.016)\,\alpha_s\, \right]
\nonumber\\
&&\Gamma(b\to c e\bar\nu_e) =
\Gamma_{0}(b\to c e\bar\nu_e) \,
\left[\, 1 + ( -0.545\pm 0.025)\,\alpha_s\, \right]
\nonumber\\
&&R=BR(b\to\tau X)/BR(b\to eX) = R_0
\left[\, 1 + ( 0.094\mp 0.009)\,\alpha_s\, \right]
\nonumber\\
&&{\rm P} = -(0.293\mp 0.002)
\left[\, 1 + ( 0.140\mp 0.015)\,\alpha_s\, \right]
\nonumber\\
&&<\, t \,>(b\to\tau X)= (0.34\pm 0.03)
\left[\, 1 + ( 0.016\pm 0.003)\,\alpha_s\, \right]
\nonumber\\
&&<\, t \,>(b\to e X)= (0.20\pm 0.02)
\left[\, 1 + ( 0.035\pm 0.007)\,\alpha_s\, \right]
\nonumber
\end{eqnarray}
where $<\, t \,>(\ldots)$ denote
the average values of $t$
for the corresponding decay chanels.
It is evident that the ${\cal O}(\alpha_s)$
corrections practically cancel
in the ratio $R$ of the branching ratios as well as
in the result for the polarization $P$.
The moments $<\, t \,>(b\to\tau X)$ and $<\, t \,>(b\to e X)$
are also insensitive to $\alpha_s$
corrections. On the other hand all these quantities are
sensitive to the quark masses and therefore they may be
used for fixing $m_b$ and $m_c$.

\section*{Acknowledgements}
MJ would like to thank Kostya Chetyrkin for helpful discussions.
This work was supported in part
by KBN grant 2P30225206, by DFG contract 436POL173193S
and by Graduierten\-kol\-leg Elementar\-teilchen\-physik at
the University of Karlsruhe.

\pagebreak[4]

\end{document}